# Realization of Zero-Refractive-Index Lens with Ultralow Spherical Aberration

Xin-Tao He [1,3], Zhi-Zhen Huang [1,2,4], Ming-Li, Chang [1,3], Shao-Zeng Xu [1,2], Fu-Li Zhao[1,3], Shao-Zhi Deng [1,2,4], Jun-Cong She [1,2,4,5,†], and Jian-Wen Dong [1,3,*]

1. State Key Laboratory of Optoelectronic Materials and Technologies, Sun Yat-sen University, Guangzhou 510275, China
2. Guangdong Province Key Laboratory of Display Material and Technology, Sun Yat-sen University, Guangzhou 510275, China
3. School of Physics, Sun Yat-sen University, Guangzhou 510275, China
4. School of Electronics and Information Technology, Sun Yat-sen University, Guangzhou 510275, China
5. Sun Yat-sen University-Carnegie Mellon University (SYSU-CMU) Shunde International Joint Research Institute, Shunde 528300, People's Republic of China

**Optical complex materials offer unprecedented opportunity to engineer fundamental band dispersion which enables novel optoelectronic functionality and devices. Exploration of photonic Dirac cone at the center of momentum space has inspired an exceptional characteristic of zero-index, which is similar to zero effective mass in fermionic Dirac systems. Such all-dielectric zero-index photonic crystals provide an in-plane mechanism such that the energy of the propagating waves can be well confined along the chip direction. A straightforward example is to achieve the anomalous focusing effect without longitudinal spherical aberration, when the size of zero-index lens is large enough. Here, we designed and fabricated a prototype of zero-refractive-index lens by comprising large-area silicon nanopillar array with plane-concave profile. Near-zero refractive index was quantitatively measured near 1.55 μm through anomalous focusing effect, predictable by effective medium**

**theory. The zero-index lens was also demonstrated to perform ultralow longitudinal spherical aberration. Such IC compatible device provides a new route to integrate all-silicon zero-index materials into optical communication, sensing, and modulation, and to study fundamental physics on the emergent fields of topological photonics and valley photonics.**

Dirac cones in fermionic systems have attracted tremendous attention in the fields of topological insulator and graphene[1-3]. Following the pace of condensed matter, these conical dispersion bands have been extended to bosonic systems particularly for electromagnetic waves[4-12]. Bosonic Dirac cones at the zone boundary reveal many similar phenomena with fermionic particles. For example, bianisotropic metamaterials can access to modulate the spin flow of light with backscattering immune at the boundary of topological photonic crystals, after opening a nontrivial gap from Dirac cone[13,8,11]. An alternative method is proposed to implement photonic analogue of the integer quantum Hall effect by using periodic coupling resonators on a silicon-on-insulator platform in near-infared (NIR) wavelength scale[7,14,15].

Beyond those predominant behaviors, bosonic Dirac cones also present extra features other than femionic systems. Recently, another type of photonic Dirac cones induced by accidental degeneracy at the zone center has been found in a class of all-dielectric photonic crystals[16,17], in which the optical response shows very different to the case at the zone boundary. One of the remarkable properties is zero-index behavior such that the effective permittivity and permeability are simultaneously to be zero at Dirac frequency. To date, nanorod-array structure is the exclusive way to realize all-dielectric zero-index photonic crystal in optical frequency regime. How to sufficiently confine the propagation wave in the plane of periodicity is a practical challenge for rod-slab structures. The first implementation at near-infrared (NIR) wavelength has been fabricated by alternating silicon/silica layers to mimic infinite rods, a smart and highly-customized method[18]. Another metal-dielectric-metal strategy has been proposed afterwards for the on-chip realization of zero-index photonic crystal in a small-area sample size[19]. As the all-dielectric zero-index response is based on the collective effects (e.g. Bragg scattering) of photonic crystals, a well-performed device is in need of large-area size so as to functionalize the zero-index properties. This is why we prefer large-area device.

Recent developments of metasurface-based flat lens have revealed many outstanding capabilities to manipulate the phase of light in micro-nano scale by using optical resonators with discrete phase distribution[20-26]. It is concentrated on out-of-plane operation that the propagating waves are vertical to the chip. Alternatively, zero-index photonic crystals and metamaterials are capable to reshape the wavefronts along the in-plane direction that is parallel to the substrate[27-33]. There are many efforts to well manipulate the phase of light such as waveguide cloaking effect[16,31], directional emitters[18,32] and unidirectional transmission[31,33], etc.

In this paper, we experimentally designed and realized a zero-refractive-index lens with plane-concave profile to achieve low-aberration device. The device was fabricated in large-area footprint by comprising of wavelength-order-thickness Si nanopillars arranged in square lattice. Anomalous focusing effect has been uniquely observed through the out-of-plane scattering from the irregular substrate at near-infrared wavelength. The focusing spot was close to the curvature center of the plane-concave lens, indicating a little phase shift change of the near zero-index photonic crystal. The effective refractive index evaluated from optical microscope images was well consistent with the retrieval of effective medium theory. Such large-size device was also demonstrated to perform ultra-low longitudinal spherical aberration.

Consider a collimated beam passing through an optical plane-concave lens in air background. General speaking, the optical plane-concave lens is made of glass with the refractive index of $n > 1$. In other words, it is optically dense medium surrounded by less-dense medium, resulting that the exit beam turns to divergence (Figure 1a). The exit beam can also become convergent by just altering the positions of the two media, e.g. air lens surrounded by glass background, however, it is not straightforward for large-area integrate devices, and the position of focal point is limited by the refractive index. The focal point is always on the right side of the concave surface center (the intersection of dash lines in Figure 1a) with certain distance. Nature optical materials do not have the capacity to achieve extreme refraction such as at-center-focusing and left-side-focusing. An alternative way is to use the concave lens constructed by low-index materials of $n < 1$ in air background. In this way, the focal point will be located at arbitrary location on both sides of the curvature center of the optical plane-concave lens. For example, the focal point is on the right of the concave center when $n < 1$, and on the left when $n < 0$ (Figure 1c), illustrating the flexibility of the on-demand design. In addition, Figure 1 also shows that the focal points of the off-axial light (purple in Figure 1) are deviated from those of the paraxial ones (yellow) if $n \neq 0$, leading to longitudinal spherical aberration (LSA), yielding

$$LSA = 10 \lg \left| \frac{\Delta_S - \Delta_L}{R} \right|, \qquad (1)$$

where such aberration is quantified in dB scale. $R$ is the curvature of optical plane-concave lens. $\Delta_S$ is the distance from the curvature center of the lens to the focal point of the off-axial light described by Snell's law, while $\Delta_L$ is related to the paraxial light described by Lens' maker formula (see Supporting Information Figure S1 for analytical derivation of LSA). To vanish LSA, it is possible to construct the optical lens by near-zero refractive index. Figure 1d illustrates a special case of the LSA-free concave lens with zero-index material. All of the refracted rays focus on the curvature center, no matter whether it is off-axial or paraxial. On the other hand, it may inspire to determine the effective refractive index of the material with plane-concave profile. By simply deduction from Supporting Information Appendix A, the effective refractive index obeys the following expression,

$$n_{eff} = \frac{R}{H}\sin\left[\arcsin\left(\frac{H}{R}\right) - \arctan\left(\frac{H}{\sqrt{R^2 - H^2} + \Delta_s}\right)\right], \quad (2)$$

where $H$ is the off-axial distance from the incident beam to the optical axis. In other words, one can experimentally retrieve the effective refractive index according to the distance $\Delta_S$ by measuring the position of focal point.

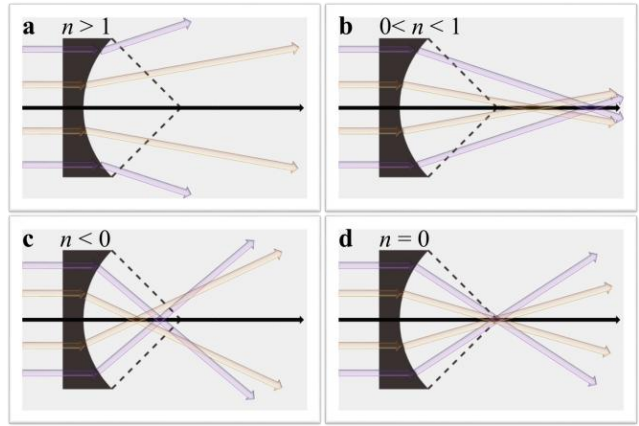

**Figure 1.** Schematic diagrams of light propagation behaviors through a concave lens. (a) Ray trace in traditional concave lens with high-index material ($n > 1$), i.e. optically dense medium surrounded by less-dense medium. The exit beam turns to become divergent. (b-c) Ray trace of light focusing effect in concave lens with positive-low-index medium ($0 < n < 1$) and negative-index medium ($n < 0$), respectively. But the focal spots illuminated by off-axial light (purple ray) are always deviated from those of the paraxial cases (yellow ray), leading to longitudinal spherical aberration (LSA). (d) LSA-free optical lens with zero-index medium ($n = 0$). In this case, all of the refracted rays focus on the concave center, no matter whether it is off-axial or paraxial.

The optical low-LSA lens can be realized by all-dielectric photonic crystal with near-zero refractive index in micro-scale. When the effective refractive index of the optical lens is close to zero, the collimated beam will be convergent with low LSA near the curvature center after propagating through the optical lens. Figure 2a shows the design of the zero-index lens with silicon nanopillars. As the propagating waves operate in the plane of chip, it is compatible to integrate the low-LSA lens with other photonic devices. Note that the nanopillars are arranged in square lattice, some of which are removed to form the concave profile. The concave surface degree is 120º and the concave surface radius is 49 μm. The overall footprint is about 49×132 μm$^2$, large enough to ensure the device serving as a well-behaved effective medium. Figure 2b gives the details of silicon rods with the refractive index of 3.42, the lattice constant of 818 nm and the rod diameter of 335 nm. The bulk band structure corresponding to zero-index photonic crystal with infinite height is plotted in Figure 2c, calculated by plane wave expansion method for transverse magnetic (TM) polarization[34]. Photonic Dirac cone can be observed at the center of the Brillouin zone where the two branches (red and blue) with conical dispersion intersect a flat quasi-longitudinal band (green) at the Dirac wavelength of λ = 1490 nm. Such conical dispersion results from the triply accidental degeneracy of the twofold degenerate modes and the non-degenerate mode at Γ point. Furthermore, the triply degenerate states are derived from monopole and dipole excitations, indicating the bulk effective indices of the device can be retrieved by the effective medium theory (EMT) based on Mie resonance[36], as shown in Figure 2d. Both bulk permittivity (blue solid) and permeability (purple dash) are equal to zero simultaneously at the triply-degenerate Dirac wavelength of 1490nm. It should be noted that a broadband single-mode region with low wavevector component is observed below the Dirac point, where is mapped as double-zero-index behavior in the practical system. On the other hand, conical dispersion still exists upon the Dirac wavelength, but it cannot be regarded as near-zero refractive index due to the excitation of quasi-longitudinal band with high wavevector component. This is also verified in the projected band diagram of Figure 2e such that the bulk states (blue shaded) are allowed in the zero-index photonic crystals.

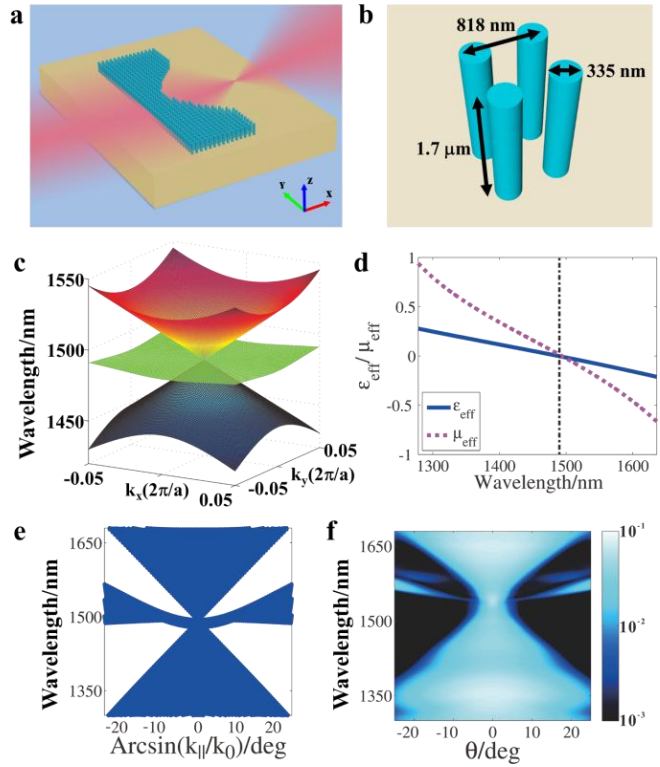

**Figure 2.** Design of optical zero-index lens on all-silicon platform. (a) Schematic view of the design of zero-index lens on silicon substrate, consisting of nanopillars in air background. When the effective refractive index of the lens is close to zero, the collimated beam will be convergent near the center of concave surface with ultralow LSA. (b) Magnified view for the unit cell of zero-index lens. In order to retrieve zero-index behavior, the pillars are arranged in a square lattice with lattice constant $a$ = 818 nm, diameter $d$ = 335 nm and height $h$ = 1.7 μm. (c) Bulk band structure of zero-index photonic crystal with infinite height. The parameters of lattice constant and diameter are the same as the above design. Photonic Dirac cone can be observed near $\Gamma$ point where the two branches (red and blue) with conical dispersion intersect a flat quasi-longitudinal band (green). Such conical dispersion results from the triply accidental degeneracy at the Dirac wavelength of $\lambda$ = 1490nm. (d) Corresponding effective permittivity $\varepsilon_{eff}$ (blue solid line) and permeability $\mu_{eff}$ (purple dash line) as a function of wavelength, obtained by effective medium theory. Note that these two parameters approach to zero simultaneously at Dirac wavelength ($\lambda$ = 1490 nm). (e) The projected band structure along $\Gamma X$ direction. The blue shaded region indicates bulk states allowed in this zero-index photonic crystal. (f) Transmission spectra of the Dirac lens with wavelength-order-thickness pillars. The conical shape spectra are well consistent with the 2D projected band of (e), indicating that the optical response of the 1.7μm-height planar system is almost equivalent to that of the 2D zero-index photonic crystal. All of the refractive index for silicon in this work is set to be 3.42.

The optical properties of the zero-index lens will deviate from those predicted in 2D model as the height of the realistic nanopillars is finite along out-of-plane direction. However, such deviation may be recovered when the nanopillars are high enough. In order to better understand this issue, we set the height to be on the order of wavelength (1.7 μm) and utilize three dimensional finite-difference time-domain (FDTD) simulations[35] to compute the transmittance of zero-index photonic crystal (Figure 2f, see Supporting Information Appendix G for the setup of numerical simulations). The conical shape transmission spectra in ($\lambda$, $\theta$) plane are well consistent with the 2D projected band diagram in Figure 2e, indicating that the optical response of the wavelength-order-thickness planar system is almost equivalent to that of the 2D zero-index photonic crystal. Note that there is no Fabry-Perot oscillation in the transmission spectra due to the suppression of out-of-plane radiative loss. This inevitable loss will reduce the transmittance to be around 10%, but it is enough to be detected in our experimental setup.

The zero-index lens was fabricated by a top-down method[37]. The Si nanopillars were etched from n-type (100) single crystalline Si substrate. Firstly, chrome (Cr) nano-masks (384 nm in diameter) were defined on the Si substrate by electron beam lithography followed by chlorine ($Cl_2$) plasma etching. Secondly, with the protection of the Cr nano-masks, Si pillars were fabricated by subsequent $SF_6$ (15 sccm)/$C_4F_8$ (65 sccm) plasma etching. The Cr-mask possesses not only high resistance to ion bombardment but also a good chemical etching selectivity to that of Si. Moreover, the introducing of $C_4F_8$ can bring a sidewall passivation to prevent the undercut etching. As a result, a well isotropic etching effect was obtained, which contributed to the forming of the high aspect ratio Si pillars. Finally, the Si pillars were thermal oxidized in 900 °C with $O_2$ (900 sccm) for 60 minutes followed by removal of surface oxide using hydrofluoric (HF) acid. With this crucial step, a clean and smooth surface was obtained, while the re-deposition product and the Cr residue were fully eliminated. This residue-free Si nano-pillar array is beneficial in PC application.

Figure 3a depicts the scanning electron microscopy (SEM) image of the fabricated zero-index lens consisting of 60-by-161 silicon nanopillars. The morphology details are shown in the insets of Figure 3a with a magnified view. The Si nanopillars have the typical height on the order of wavelength (1.7 μm), guaranteeing the nanopillars long enough to confine sufficient energy along the in-plane propagating direction. Note that the shapes of nanopillars are cone instead of cylinder. The diameter of the pillar is typically 310 nm at the top while 400 nm at the base, deviated from the ideal radius of 335 nm. As the photonic Dirac cone at the center of momentum space results from the accidental degeneracy, it is likely to suffer from a mini gap. Such cone-shaped pillars will be beneficial to eliminate the mini gap so as to obtain the gapless transmission, although the total energy has one order of magnitude less than ideal. On the other hand, this imperfection impacts a little on the near-zero-index behavior near the Dirac wavelength, even it has been taken into account in the effective medium theory (see more details in Supporting Information Appendix B).

In order to experimentally demonstrate the zero-index feature of the fabricated lens, the optical microscope images were captured, recording the light propagation through the device (see Supporting Information Figure S3 for measure setup). The experimental incoherent focusing images are shown in Figures 3b-e at different operation wavelength (see Supporting Information Appendix D). White solid lines indicate the location of the zero-index concave lens and the intersection of the two dash lines is the curvature center of the concave profile. The light trajectory came from the scattering signal that passes through the lens and reflects by irregular silicon substrate. By fine tuning the incident spot along the y direction, the direction of the scattering signal changed but it is always convergent to the middle in Figures 3b-d, near the curvature center of the lens. Such anomalous focusing effects are relative to the low refractive index of $n_{eff} < 1$, which is absent in conventional concave lens (quantitative demonstration will be discussed later). We also observed the evolution of the focal spots along optical axis at the three different operation wavelengths. The spot is first on the right side of the curvature center at the shorter wavelength of 1430 nm (Figure 3b), and then near the center at the Dirac wavelength of 1490 nm (Figure 3c), finally on the left side at the longer wavelength of 1550 nm (Figure 3d). But the focusing effect will be difficult to be observed when the wavelength increases to 1610 nm (Figure 3e). To completely demonstrate the evolution of the focal points, we plot the intensity of focusing pattern as a function of wavelength and position, depicted in Figure 3f. The focal spots continuously experience from the right to left side of the center of concave surface. In the long-wavelength region (above 1580nm), an additional bright region is located on the top-rightmost, increasing the difficulties of focusing measurement. This is due to the excitation of quasi-longitudinal mode. In comparison, we also performed a control experiment to show the divergent behaviors in homogenous silicon concave lens (see Supporting Information Appendix F).

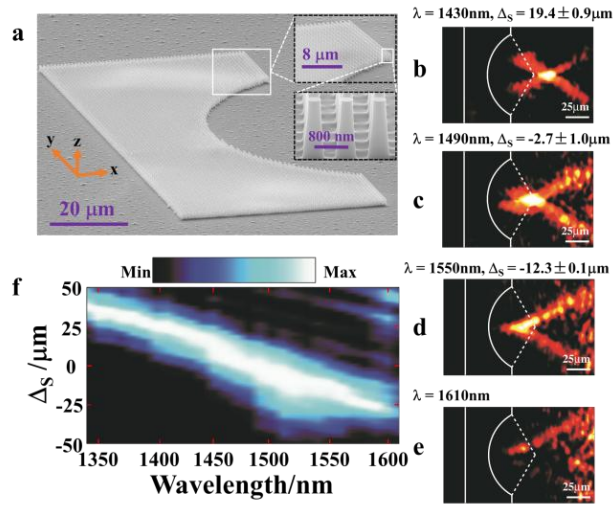

**Figure 3.** Fabricated device and optical characterization. (a) SEM image of the fabricated zero-index lens. The large-area device consist of 60-by-161 silicon nanopillars, some of which are removed to take concave shape. More details are shown in the insets. The Si nanopillars have the typical height of 1.7 μm, guaranteeing the nanopillars long enough to confine sufficient energy along the in-plane propagating direction. (b-e) Optical microscope images of two incoherent beams passing through the zero-index lens. We can observe (b) right-center-focusing at 1430 nm, (c) near-center-focusing at 1490 nm and (d) left-center-focusing at 1550 nm, respectively. Such anomalous focusing effects furnish direct evidences for the effective near-zero refractive index of the fabricated lens. Note that the focusing effect will be difficult to be observed when the wavelength increases to (e) 1610 nm. (f) Intensity of focusing pattern measured from off-axial illumination, as a function of wavelength and position. The focal spots continuously experience from the right to left side of the center of concave surface. In the long-wavelength region (above 1580nm), an additional bright spot is on top-rightmost increasing the difficulties of focusing measurement, due to the excitation of quasi-longitudinal mode.

Figure 4 illustrates how performance of device can be used to demonstrate the functionality of the fabricated lens. Figure 4a plots the quantitative value of the effective refractive index as a function of wavelength. According to eq 2, the index can be evaluated from the distance $\Delta_s$ of the focal spot and the curvature center of the optical lens. By carefully estimation stated in Supporting Information Appendix E, the effective refractive index by experimental evaluation results are then plotted in blue in Figure 4a. With tuning the operation wavelength, one can see that the effective refractive index experiences from positive ($0.417 \pm 0.002$ at 1340 nm) to negative refraction ($-0.505 \pm 0.014$ at 1570 nm), which is good agreement with the trends of effective medium theory (red dash curve in Figure 4a). As predicted by the band theory in Figure 2, we find the effective refractive index is very close to zero ($-0.047 \pm 0.017$) at the Dirac wavelength of 1490 nm. Such zero-index concave lens will have low longitudinal spherical aberration caused by the difference between the focal spots of the paraxial and the off-axial light. We also plot the value of LSA in both theory and experiment in Figure 4b. See Supporting Information Appendix E for more details. At the Dirac wavelength of 1490 nm, the measured LSA decreased to $-28.3 \pm 3.7$ dB, at least 10 dB less than the other wavelengths. Such dip in LSA spectrum is relative to zero- index behaviors. These results verify quantitatively that we have obtained an ultralow-LSA zero-index lens. But in the long-wavelength region (above 1510nm), the metalens will be deviated from near-zero index system since the flat quasi-longitudinal state is more likely to be excited, increasing the uncertainty of focal spots and LSA (see errorbars in Figures 4a-4b). The anomalous focusing effect will be difficult to be observed when the wavelength increases to upon 1580 nm.

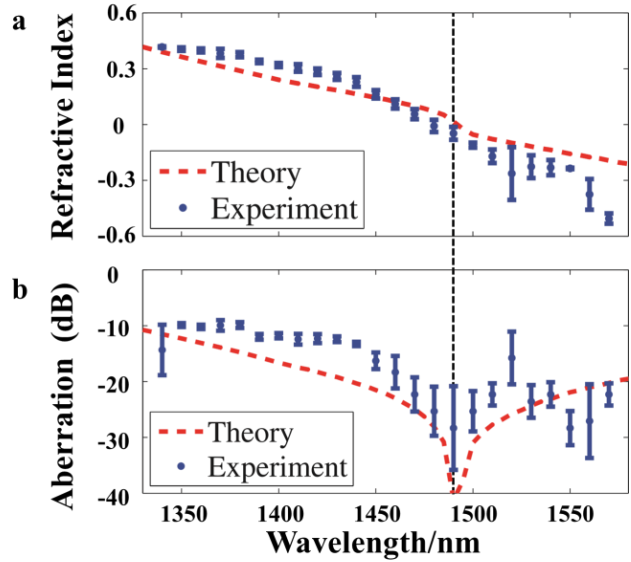

**Figure 4.** Quantification of device performance with near-zero index. (a) Effective refractive index and (b) longitudinal spherical aberration of the zero-index lens as a function of wavelength. The red line is retrieved by effective medium theory, while the blue points are experimentally evaluated by focusing images. Error bars represent the uncertainty of focal spots. In the wavelength region from 1340 to 1570 nm, the effective refractive index experienced from positive, zero to negative refraction, which is good agreement with the trends of effective medium theory. At the Dirac wavelength of 1490 nm (black dash), the measured LSA decreased to $-28.3\pm3.7$ dB, at least 10 dB less than the other wavelengths. Such dip in LSA spectrum is relative to zero-index behaviors. But the optical lens cannot be mapped to near-zero index system in the long-wavelength region (above 1510nm), since the flat quasi-longitudinal state is likely to be excited. It will increase the uncertainty of focal spots such that the anomalous focusing effect almost vanishes when the wavelength is over 1580 nm.

In summary, we have experimentally demonstrated the all-dielectric zero-index photonic crystal on silicon chip, which is functionalized by a plane-concave lens with photonic Dirac cone at near IR wavelength. Such zero-index lens was large-area fabricated on all-silicon platform and was optically characterized according to anomalous focusing effect. The zero-index feature was quantitatively verified near Dirac wavelength to retrieve ultralow longitudinal spherical aberration.

Our IC compatible strategy provides a route to integrate all-silicon zero-index materials into optical communication, sensing, and modulation, and to study fundamental physics on the emergent fields of topological photonics and valley photonics.